\documentclass[12pt]{iopart}
\usepackage{epsfig,graphicx,dcolumn,bm}

\begin{document}

\title{Dynamic analysis of influential stocks based on conserved networks}

\author{Xin-Jian Xu$^{1}$, Min Qin$^{1}$, Xiao-Ying Song$^{2}$ and Li-Jie Zhang$^{3}$}

\address{$^{1}$Department of Mathematics, Shanghai University, Shanghai 200444, People's Republic of China\\
$^{2}$School of Economics, Shanghai University, Shanghai 200444, People's Republic of China\\
$^{3}$Department of Physics, Shanghai University, Shanghai 200444, People's Republic of China}
\ead{lijzhang@shu.edu.cn}

\begin{abstract}
Characterizing temporal evolution of stock markets is a fundamental and challenging problem. The literature on analyzing the dynamics of the markets has focused so far on macro measures with less predictive power. This paper addresses this issue from a micro point of view. Given an investigating period, a series of stock networks are constructed first by the moving-window method and the significance test of stock correlations. Then, several conserved networks are generated to extract different backbones of the market under different states. Finally, influential stocks and corresponding sectors are identified from each conserved network, based on which the longitudinal analysis is performed to describe the evolution of the market. The application of the above procedure to stocks belonging to Standard \& Pool's 500 Index from January 2006 to April 2010 recovers the 2008 financial crisis from the evolutionary perspective.
\end{abstract}


\maketitle

\section{Introduction}
Stock markets are well-defined complex systems consisting of multi heterogeneous stocks with complex relationships among them~\cite{mantegna00book}. The prices of the stocks evolve as a consequence of their internal and external interactions, and different assets present turbulent financial time series making the market behaviors even more difficult to be examined. Therefore, it is important to mine essential information from the market and build an efficient model to characterize its dynamic properties, which not only provides a fundamental understanding of financial systems but also provides practical insights for policymakers and practitioners~\cite{campbell97book}.

One seminal approach is the random matrix theory~\cite{laloux9prl,plerou99prl,potters21book} which characterizes the eigenvalue distribution of the correlation coefficient matrix of time series of stocks and have unveiled many stylized facts of stock markets~\cite{eom09phya,song11pre,jiang14sc,wang17chaos}. For instance, a stock market containing many business sectors (groups of stocks sharing common economic properties) with hierarchial organization~\cite{djauhari16jsm}. However, the random matrix could not draw interactions well among these sectors. To understand the market in a more exact way, the complex network theory~\cite{newman03book} was adopted instead. Examples include the minimal spanning tree~\cite{bonanno03pre}, the asset graph~\cite{onnela04epjb}, the planar maximally filtered graph~\cite{tumminello05pnas} and the threshold network~\cite{boginski05csda}. Based on these models, many topological characteristics have been observed for the markets such as New York Stock Exchange~\cite{tumminello07epjb,liu11qf,wang17jeic,xu18phya}, German Stock Exchange~\cite{wilinski13phya}, Tokyo Stock Exchange~\cite{wilinski18jsm}, Hong Kong Stock Market~\cite{xu17sc} and Shanghai Stock Market~\cite{yang13mplb}.

In general, a stock market evolves with economic states, resulting in sequential changes of stock prices from one state to another~\cite{musciotto18pc}. To explain the development of the economic state from the perspective of the market, there is an increasing interest in characterizing temporal evolution of stock networks. Yet, most studies focused on the evolution of global topologies, such as the edge density, the average clustering coefficient and the average shortest path length, which only provides the macro topological information of the market corresponding to different states~\cite{xu18phya,xu17sc}. To get a deeper understanding of the dynamics of the market, it is essential to study the evolution of the market from a micro point of view~\cite{bardoscia12jsm,bongiorno21epl}. Specially, how to identify influential stocks and evaluate their roles in the diffusion of microfinance is of great importance~\cite{abhijit13sci}.

There are two major approaches to identify influential stocks in the literature. The first approach is from the dynamical point of view. For instance, Wang et al.~\cite{wang16epjb} and Benzaquen et al.~\cite{benzaquen17jsm} suggested the cross-response (impact) function to characterize the influence of a stock. That is, the stock with strong cross-response has large influence on other stocks, resulting in the rank of stocks. The second approach is from the structural point of view. Specially, the concept of centrality has been widely used to rank nodal influence. For example, Roy and Sarkar~\cite{roy11proc} employed the degree centrality to rank stocks. They compared top $10$ influential stocks corresponding to pre- and post-crisis and observed that the change in ranks of top $3$ influential stocks are relatively low compared to those ranked lower. Nevertheless, a subtle analysis remains demanding.

The goal of this paper is to identify most influential stocks of a stock market from the evolutionary perspective such that it can recover a financial crisis efficiently. To this end, we first construct a series of threshold networks for stocks in an investigating period. Then, we consider different stages of the crisis and build a conserved network for each stage. Finally, we identify influential stocks and corresponding sectors to describe crisis propagation. To test its efficacy, we apply our framework to stocks belonging to Standard \& Pool's (S\&P) 500 Index from January 2006 to April 2010 and recovers the 2008 financial crisis in an evolutionary way.

\section{Methodology}\label{sec2}
Section~\ref{sn} introduces the Pearson correlation coefficient, the P-threshold method with multiple hypothesis testing and the moving-window method to construct the dynamic sequence of stock networks. Section~\ref{cn} presents conserved networks associated with different stages of a financial crisis. Section~\ref{nc} introduces four typical centralities to measure nodal influence. Section~\ref{os} presents the order statistic to synthesize centralities of nodes to rank their influence.

\subsection{Stock networks}\label{sn}
Let $p_i(\tau)$ $(i=1,2,\cdots,N; \tau=1,2,\cdots,M)$ be the daily closing price of stock $i$ at time $\tau$, one obtains the logarithmic return of $i$ over a time interval $\Delta\tau$ by
\begin{equation}\label{return}
r_i(\tau)=\ln p_i(\tau)-\ln p_i(\tau-\Delta\tau).
\end{equation}
In this paper we set $\Delta\tau=1$, so $r_i(\tau)$ represents the daily return of stock $i$ at time $\tau$. Then, the correlation coefficient between stocks $i$ and $j$ is defined by
\begin{equation}\label{correlate}
w_{ij}=\frac{\langle r_i r_j\rangle - \langle r_i\rangle \langle r_j\rangle }{\sigma_i \sigma_j},
\end{equation}
where $\langle r_i \rangle=\sum_{\tau=1}^{M} r_i(\tau)/M$ is the mean and $\sigma_i=\sqrt{\sum_{\tau=1}^{M}\left[ r_i-\langle r_i \rangle\right]^2/M}$ is the standard deviation. The ensemble of $w_{ij}$ forms the correlation matrix $\bm{W}$ of a stock market in a window of width $M$.

To filter $w_{ij}$, we use the P-threshold method and set the following hypothesis test~\cite{xu17sc},
\begin{eqnarray}
H_0: & w_{ij}=0,\\
H_1: & w_{ij}\neq 0.
\end{eqnarray}
The corresponding test statistic is
\begin{equation}
T_{ij}=w_{ij}\sqrt{\frac{n-2}{1-w_{ij}^2}} \sim t_{n-2},
\end{equation}
where $n$ is the sample size and $n-2$ is the degree of freedom. Given a significance level $\alpha$, one should reject $H_0$ if the absolute value of the test statistic exceeds the cut-off value $t_{\alpha/2}(n-2)$, namely, if
\begin{equation}
|T_{ij}|>t_{\alpha/2}(n-2).
\end{equation}
To maximize the number of discoveries while controlling the fraction of false discoveries, we perform the multiple hypothesis test based on the Bonferroni correction. Specially, for the significance level of $\alpha=0.01$, we include any interactions between stocks if $|T_{ij}|>t_{\alpha/N(N-1)}(n-2)$. Since the Bonferroni correction assumes complete independence between the tested p-values, one may consider further the False Discovery Rate (FDR) approach~\cite{fdr} to relax the assumption of independence. As a consequence, more edges among stocks will be maintained. For smoothing purpose, we adopt the moving-window method~\cite{onnela03pre}. Assuming the width of each window is $M$ and the sliding interval is $\Delta M$, one can obtain a series of windows overlap with each other for any oberving period with proper choices of $M$ and $\Delta M$. Inside each window, an edge is created between a pair of stocks $i$ and $j$ if $w_{ij}\neq 0$. This process is repeated throughout all the elements of the correlation matrix and finally a stock network is generated. Specially, we assume the network $G(V,E)$ is unweighted and undirected, which can be described by an adjacency matrix $\bm{A}=(a_{uv})_{N \times N}$ with elements
\begin{equation}\label{matrix}
a_{uv}=\biggl\lbrace
\begin{array}{ll}
1, & \quad \mbox{if $u$ and $v$ are connected,}\\
0, & \quad \mbox{otherwise.}
\end{array}
\end{equation}

\subsection{Conserved networks}\label{cn}
A typical stock market usually experiences various financial situations, including bull and bear runs, business as usual and financial crises. Of great importance is to delve into reliable indicators of the crisis
from the market. However, the 2008 financial crisis has highlighted the main limitations of standard models, as they cannot detect the crisis even by using posterior data~\cite{stiglitz16book}. Here, we address this issue by means of conserved networks. According to different states associated with a crisis, we divide the whole investigating period into $5$ stages: the normal stage before the crisis, the stage of the transition from the normal state to the crisis, the stage during the crisis, the stage of the transition from the crisis to the normal state and the stage after the crisis. Inside any stage, there are a number $K$ of consecutive windows overlap with each other, based on which $K$ stock networks are constructed. Furthermore, we assume that the interactions between significant stocks will persist while the interactions between insignificant stocks will vary with time, which leads to the idea of conserved networks: for any pair of stocks, an edge between them in the corresponding conserved network exists if and only if all these $K$ networks within the stage have this edge. As a consequence, we obtain $5$ conserved networks, which characterize dynamic characteristics of the investigating period.

\subsection{Centrality measures}\label{nc}
The most influential stocks may help us understand risk propagation in a stock market and design corresponding control measures. To represent nodal influence in each conserved network, we adopt the concept of centrality. In this paper, we consider four typical measures: degree centrality (DC), eigenvector centrality (EC), closeness centrality (CC) and betweenness centrality (BC).

Given a stock network $G(V,E)$, the DC of node $u \in V$ is define by~\cite{freeman78sn}
\begin{equation}\label{dc}
\texttt{DC}(u)=k_{u},
\end{equation}
where $k_{u}=\sum_{v=1}^{|V|}a_{uv}$ is the degree of node $u$ and $|V|$ is the number of nodes. Although the DC is the simplest centrality measure, it can be illuminating. In stock markets, it seems reasonable to suppose that stocks with connections to many others might have more access to information than those with fewer connections. A natural extension of the DC is EC, defined as~\cite{bonacich87ajs}
\begin{equation}
\texttt{EC}(u)=\bm{\upsilon}_u^{\texttt{max}},
\end{equation}
where $\bm{\upsilon}^{\texttt{max}}$ is the eigenvector corresponding to the largest eigenvalue of the adjacency matrix $\bm{A}$ and $\bm{\upsilon}_u^{\texttt{max}}$ is the $u$th element of $\bm{\upsilon}^{\texttt{max}}$ corresponding to stock $u$. As a consequence, the stock $u$ with larger $\texttt{EC}(u)$ can be important because it has many neighbors (even though those stocks may not be important themselves) or because it has important neighbors of high degrees.

Both the DC and EC only consider local information of a network. Regarding global information, some entirely different measures of centrality have been suggested incorporating shortest path lengths. One is CC, defined as~\cite{freeman78sn}
\begin{equation}\label{cc}
\texttt{CC}(u)=\frac{|V|-1}{\sum_{v \in V}l(u,v)},
\end{equation}
where $l(u,v)$ is the shortest path length from node $u$ to node $v$. According to Eq.~(\ref{cc}), the smaller average distance from the stock $u$ to others, the larger value of $\texttt{CC}(u)$ it has. BC is another different concept of centrality, initially proposed by Bavelas~\cite{bavelas48ho} and generalized by Freeman~\cite{freeman77soc},
\begin{equation}
\texttt{BC}(u)=\frac{\eta(s,u,t)}{\sum_{s\neq u\neq t}\eta(s,t)},
\end{equation}
where $\eta(s,u,t)$ is the number of those shortest paths passing through $u$ and $\eta(s,t)$ is the total number of the shortest paths from node $s$ to node $t$. In contrast to the CC, the BC measures the extent to which a stock lies on paths between other stocks.

\subsection{Order statistic}\label{os}
Different centrality measures yield different ranks of nodal influence. To synthesize multiple centralities, we regard each rank as an order statistic~\cite{das10book} and obtain a Q-statistic from the joint cumulative distribution of the $n$-dimensional order statistic:
\begin{equation}
Q(\gamma_1,\gamma_2,\cdots,\gamma_n)=n!\int_0^{\gamma_1}\int_{q_1}^{\gamma_2}\cdots\int_{q_{n-1}}^{\gamma_n}\texttt{d}q_n\texttt{d}q_{n-1}\cdots\texttt{d}q_1,
\end{equation}
where $\gamma_i$ is the rank ratio for centrality $i$ and $q_{i}$ is the lower bound of the $(i+1)$th order statistic. In present work, we use $4$ centrality measures, hence $n=4$. In fact, the above integration can be calculated in a fast way
\begin{equation}
Q(\gamma_1,\gamma_2,\cdots,\gamma_n)=n!Q_n
\end{equation}
with $Q_k=\sum_{i=1}^k(-1)^{i-1}Q_{k-i}\gamma_{k-i+1}^i/i!$. The larger value of $Q$, the greater influence the node has.

\section{Application to S\&P 500 stocks}
To validate the above framework, we apply it to stocks belonging to S\&P 500 Index. The data are daily records and the investigating period ranges from January 2006 to April 2010, yielding $1089$ observations of $422$ stocks. During the period, there was an economic crisis: the 2008 financial crisis.

\begin{figure}[h]
\centering
\includegraphics[width=0.7\columnwidth]{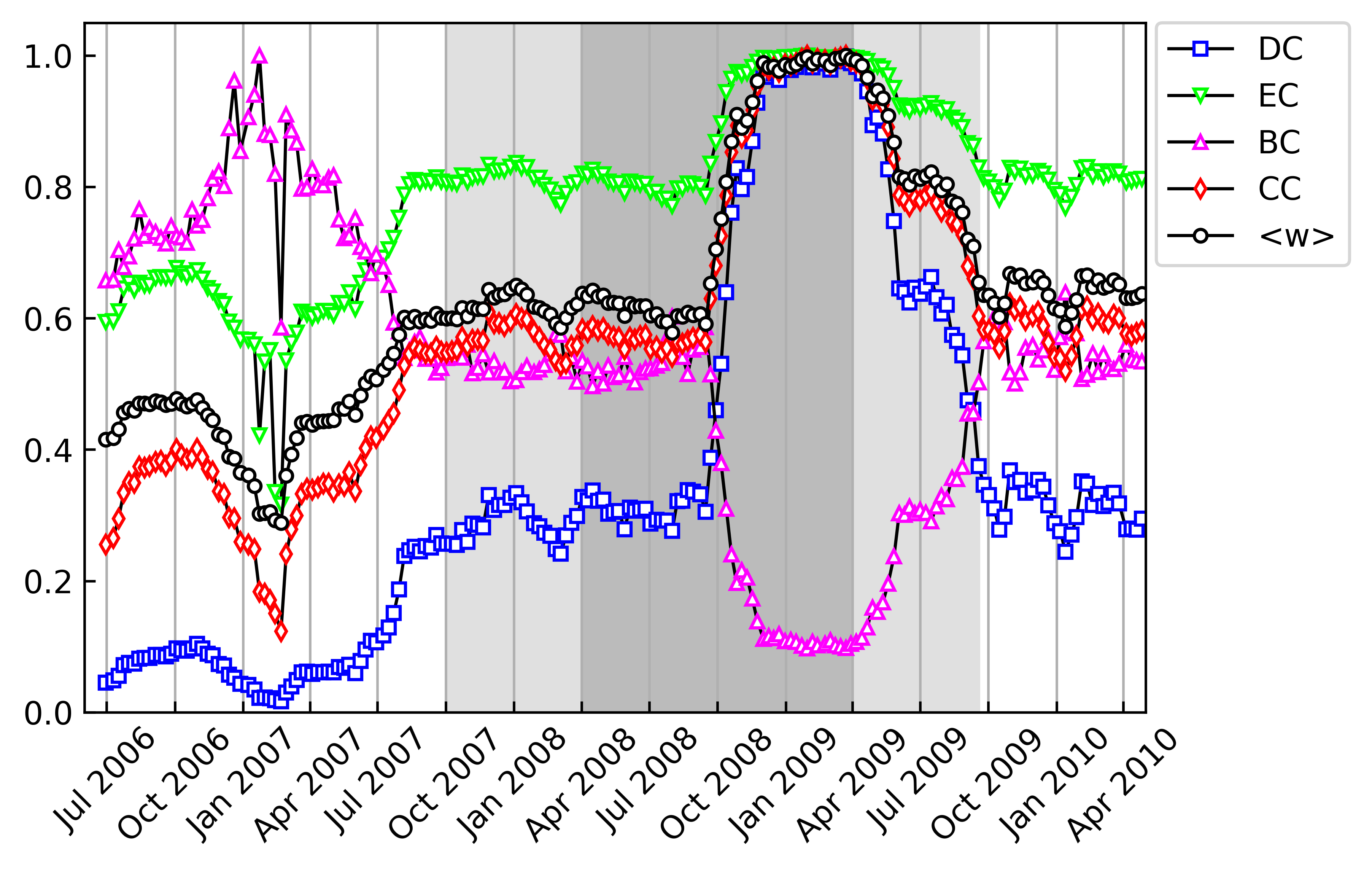}
\caption{(Color online) Temporal evolution of the normalized DC, EC, BC and CC. The average correlation coefficient $\langle w\rangle$ is presented for comparison.}\label{fig1}
\end{figure}

\subsection{Dynamic networks}\label{spsn}

First of all, we divide the investigated period through moving windows. By setting the width of each window as $M=125$ (about half a year) and the moving step as $\Delta M=5$ (about a week), we obtain $193$ windows. For each window, we fix the significance level at $\alpha=0.01$, based on which a correlation network is constructed. As a result, we obtain $193$ consecutive networks for the whole period. Then, we calculate the centrality of each node, the average of which is used to stand for the characteristics of the network. For each network, we consider the DC, EC, CC and BC, respectively. Figure 1 shows the evolution of four centralities in comparison to the evolution of the average correlation coefficient $\langle w\rangle$. For comparison, we normalize each plot by the corresponding maximum of the investigating period. However, it is not essential. The dark gray interval corresponds to the middle stage of the crisis and the two light gray intervals respectively correspond to the early transition from the normal state to the crisis and the late transition from the crisis to the normal state. Remarkably, the DC, EC and CC display the same trend with $\langle w\rangle$, while the BC evolves in the opposite way. According to Eq.~(\ref{dc}), one has $\langle \texttt{DC}\rangle=\sum_{u\in V}k_u/|V|=2|E|/|V|=(|V|-1)e$, where $|E|$ is the number of edges and $e=2|E|/|V|/(|V|-1)$ is the density of edges. Moreover, the edge density is proportional to $\langle w\rangle$. Therefore, both the DC and EC exhibit the same trend as $\langle w\rangle$. As to the CC and BC (see Eqs. (10) and (11)), the higher density of edges, the smaller distance that a node reach all the others, hence the larger value of the CC. On the contrary, the number of connected pairs of nodes increases, resulting in the decrease of the BC. Overall, the four measures can serve as good indicators of the market evolution from the macro perspective.

\begin{table}[h]
\centering
\caption{Basic statistics of $5$ conserved networks based on the Bonferroni correction.}\label{alpha1}
 \begin{tabular}{|c|c|c|c|c|c|}
 \hline
    &\textbf{$|V|$} &\textbf{$|E|$} &\textbf{$\langle k\rangle$} &\textbf{$\langle c\rangle$} &\textbf{$\langle l\rangle$}\cr\hline
    Conserved Network 1 &422 &544   &2.5782   &0.1820 &412.4460\cr\hline
    Conserved Network 2 &422 &7556  &35.8104  &0.4562 &176.1770\cr\hline
    Conserved Network 3 &422 &9493  &44.9905  &0.5363 &139.9614\cr\hline
    Conserved Network 4 &422 &24658 &116.8626 &0.6864 &52.0883\cr\hline
    Conserved Network 5 &422 &7756  &36.7583  &0.4839 &186.8250\cr\hline
 \end{tabular}
\end{table}

\subsection{Conserved networks}\label{spcn}
Considering the 2008 financial crisis, we divide the period from January 2006 to April 2010 into $5$ stages: the normal stage before the crisis, the stage of the transition from the normal state to the crisis, the stage during the crisis, the stage of the transition from the crisis to the normal state and the stage after the crisis. For each stage, we generate a conserved network. Table~\ref{alpha1} shows the basic characteristics of the $5$ conserved networks, including the number of stocks $N$, the number of edges $E$, the average degree $\langle k\rangle$, the average clustering coefficient $\langle c\rangle$ and the average shortest path length $\langle l\rangle$. One notices apparent differences between these networks. For instance, the conserved network in the crisis is relatively dense because of the higher systemic risk. As a result, $\langle k\rangle$ and $\langle c\rangle$ are larger while $\langle l\rangle$ is smaller. But these topologies only provide macro information of the market development.

\begin{table}[h]
\centering
\caption{Top $10$ influential stocks of the conserved network corresponding to the normal stage before the 2008 crisis.}\label{tabstage1}
 \begin{tabular}{|c|c|c|c|}
 \hline
    \textbf{Stock} &\textbf{Sector} &\textbf{$Q$} &\textbf{Cumulative return}\cr\hline
    KIM &Real Estate &1     &1.3755\cr\hline
    RJF &Financials  &0.9862 &1.2366\cr\hline
    ESS &Real Estate &0.9625 &1.2327\cr\hline
    REG &Real Estate &0.9625 &1.2188\cr\hline
    BXP &Real Estate &0.9625 &1.3574\cr\hline
    UDR &Real Estate &0.9625 &1.0163\cr\hline
    FRT &Real Estate &0.9534 &1.4069\cr\hline
    LNC &Financials  &0.9449 &1.2018\cr\hline
    DHR &Health Care &0.9324 &1.4872\cr\hline
    PH  &Industrials &0.9241 &1.6564\cr\hline
 \end{tabular}
\end{table}

\subsection{Influential stocks}\label{spis}
The 2008 financial crisis is due to the U.S. financial problem of subprime mortgages. Mortgage is a loan taken from bank to buy a house, which is an agreement between homebuyers and banks. In general, people with low credit and low income can not get a loan from retail banks. But since year 2000's, a third party got involved, namely investment banks. These investment banks started buying mortgage agreements from the retail banks. In this way, the retail banks sold the loans to investment banks to have zero liability and the mortgages bought by the investment banks were used to form Mortgage Backed Security. Although the Mortgage Backed Security is a special category of subprime mortgages with higher risk, it yields higher return at the same time. As more people were becoming eligible for the mortgage, the demand for homes started increasing. More people had money (borrowed) to buy a new home. The price of residential properties only went up, hence creating the bubble. Table~\ref{tabstage1} lists $10$ most influential stocks identified from the conserved network corresponding to the normal stage before the crisis (from June 30th 2006 to September 25th 2007). As is expected, $6$ stocks are from the real estate sector.

\begin{table}[h]
\centering
\caption{Top $10$ influential stocks of the conserved network corresponding to the early transition from the normal state to the crisis.}\label{tabstage2}
 \begin{tabular}{|c|c|c|c|}
 \hline
    \textbf{Stock} &\textbf{Sector} &\textbf{$Q$} &\textbf{Cumulative return}\cr\hline
    AMP  &Financials  &1     &0.8696\cr\hline
    GL   &Financials  &1     &0.9815\cr\hline
    XOM  &Energy      &0.9945 &0.9380\cr\hline
    GE   &Industrials &0.9812 &0.9034\cr\hline
    AXP  &Financials  &0.9719 &0.7710\cr\hline
    MET  &Financials  &0.9718 &0.8784\cr\hline
    BEN  &Financials  &0.9626 &0.7921\cr\hline
    DD   &Materials   &0.9534 &0.8589\cr\hline
    WY   &Real Estate &0.9443 &0.9217\cr\hline
    TROW &Financials  &0.9442 &0.9508\cr\hline
 \end{tabular}
\end{table}

Investment banks made the loans were issued to people who had little capability to payback the loan. Inevitably, some of them eventually could not afford the monthly payments and their property went for foreclosure. At a point in 2007-2008, there were more houses on sale than there were buyers for it. This triggered a steady price fall. The housing bubble burst. When property prices started going down, people who had bought the property with the sole purpose of \lq\lq buying low and selling high\rq\rq stopped paying the mortgage. This led to more loan defaults and more foreclosures. As a result, the share price of the Mortgage Backed Security started to fall continuously and eventually started to affect on big investors that could not cover this urgency. So the crisis came into being. Table~\ref{tabstage2} lists $10$ most influential stocks identified from the conserved network corresponding to the early transition from the normal state to the crisis (from October 2nd 2007 to Marc 26th 2008), among which $6$ stocks are financials, indicating the first shock of the crisis.

\begin{table}[h]
\centering
\caption{Top $10$ influential stocks of the conserved network corresponding to the stage during the crisis.}\label{tabstage3}
 \begin{tabular}{|c|c|c|c|}
 \hline
    \textbf{Stock} &\textbf{Sector} &\textbf{$Q$} &\textbf{Cumulative return}\cr\hline
    BEN  &Financials  &1     &0.5093\cr\hline
    TROW &Financials  &1     &0.5309\cr\hline
    ITW  &Industrials &0.9900 &0.6188\cr\hline
    MMM  &Industrials &0.9835 &0.6185\cr\hline
    AXP  &Financials  &0.9812 &0.2824\cr\hline
    PCAR &Industrials &0.9719 &0.5449\cr\hline
    ARE  &Real Estate &0.9718 &0.3619\cr\hline
    BXP  &Real Estate &0.9623 &0.3377\cr\hline
    DD   &Materials   &0.9603 &0.2195\cr\hline
    AMP  &Financials  &0.9534 &0.3576\cr\hline
 \end{tabular}
\end{table}

After the burst of the bubble in the housing market, many investment banks had more liabilities than assets and faced a big trouble of liquidity. For example, the New Century Financial Corporation filed for Chapter 11 bankruptcy protection in April 2008 because of repurchase agreements and the Lehman Brothers announced bankrupt in September 2008 due to asset deterioration. In addition to investment banks, many insurance companies and financial institutions have also been greatly impacted. A conspicuous example is the American International Group, which lost $250$ billion dollars in the second quarter of 2008 and was taken over by the U.S. government finally. Soon after the earthquake in the financial market, the real economy was also shocked. On the one hand, the depression of the housing market caused correlated companies to fold. One the other hand, the rising unemployment and the shrink of personal wealth decreased consuming intention for industrial products. Table~\ref{tabstage3} lists $10$ most influential stocks identified from the conserved network corresponding to the stage during the crisis (from April 2nd 2008 to March 30th 2009). We find that not only financial institutions but also correlated industrial companies were involved with enormous losses.

\begin{table}[h]
\centering
\caption{Top $10$ influential stocks of the conserved network corresponding to the late transition from the crisis to the normal state.}\label{tabstage4}
 \begin{tabular}{|c|c|c|c|}
 \hline
    \textbf{Stock} &\textbf{Sector} &\textbf{$Q$} &\textbf{Cumulative return}\cr\hline
    ETN  &Industrials &1     &1.6504\cr\hline
    PCAR &Industrials &1     &1.5776\cr\hline
    DOV  &Industrials &0.9812 &1.5289\cr\hline
    EMR  &Industrials &0.9719 &1.4723\cr\hline
    SWK  &Industrials &0.9626 &1.4621\cr\hline
    RTX  &Industrials &0.9618 &1.4746\cr\hline
    MMM  &Industrials &0.9534 &1.5338\cr\hline
    CMI  &Industrials &0.9532 &1.8777\cr\hline
    TROW &Financials  &0.9442 &1.6997\cr\hline
    WMB  &Energy      &0.9442 &1.6594\cr\hline
 \end{tabular}
\end{table}

To contain the crisis, the Troubled Assets Relief Program was carried out in October 2008, which authorized the United States Treasury to spend up to $700$ billion dollars to purchase trouble assets both domestically and internationally. The act was widely credited with restoring stability and liquidity to the financial sector, unfreezing the markets for credit and capital and lowering borrowing costs for households and businesses. This, in turn, helped restore confidence in the financial system and restart economic growth. Another dose of fiscal stimulus is monetary easing. The lower interest rate spurred businesses to make new investments, spurred industrials to invest in renovations and spurred purchases of major durable goods like cars. Table~\ref{tabstage4} lists $10$ most influential stocks identified from the conserved network corresponding to the late transition from the crisis to the normal state (from April 6th 2009 to September 18th 2009), among which $8$ stocks are industrials, implying the initial recovery.

\begin{table}[h]
\centering
\caption{Top $10$ influential stocks of the conserved network corresponding to the normal state after the crisis.}\label{tabstage5}
 \begin{tabular}{|c|c|c|c|}
 \hline
    \textbf{Stock} &\textbf{Sector} &\textbf{$Q$} &\textbf{Cumulative return}\cr\hline
    L   &Financials  &1     &1.1066\cr\hline
    EMR &Industrials &0.9994 &1.2794\cr\hline
    ALB &Materials   &0.9906 &1.2975\cr\hline
    EMN &Materials   &0.9899 &1.2640\cr\hline
    DOV &Industrials &0.9812 &1.3617\cr\hline
    VNO &Real Estate &0.9811 &1.2271\cr\hline
    WMB &Energy      &0.9718 &1.3315\cr\hline
    UNM &Financials  &0.9534 &1.1416\cr\hline
    AFL &Financials  &0.9533 &1.2802\cr\hline
    HON &Industrials &0.9438 &1.2016\cr\hline
 \end{tabular}
\end{table}

In fact, the speed of the recovery from the 2008 financial crisis has been unusually slow. Nevertheless, under the percolation of the stimulating policy of quantitative easing, the U.S. economy began to recover since the middle of 2009. With the reduction of the systematic risk and the rising opportunity for business, the stock market boomed again.  Table~\ref{tabstage5} lists $10$ most influential stocks identified from the conserved network corresponding to the stage after the crisis (from September 25th 2009 to April 26th 2010), suggesting that the market is active across various sectors, including financials, industrials, materials, real estate and energy.

\section{Discussion}

We have synthesized the DC, EC, CC and BC by means of the Q-statistic. Although it does rank stocks in each stage, the values of $Q$ are relative close. Therefore, is is nature to ask do top $10$ influential stocks vary and how much is the variance if the procedure to construct the network is changed?

To answer this question, we first consider the change in the P-value. We perform simulations for $\alpha=0.02$ and shown the corresponding results in Appendix. Comparing Tables~\ref{alpha1} and \ref{alpha2}, we find that all the basic characteristics of the generated networks are of the same order, indicating similar structure. We also compare top $10$ influential stocks in each stage as $\alpha$ increases from $0.01$ to $0.02$. Specially, the stock AVB belonging to Real Estate (Table~\ref{tabstage1a}) replaces the stock PH belonging to Industrials (Table~\ref{tabstage1}) in Stage 1; the stock GPC belonging to Consumer Discretionary (Table~\ref{tabstage2a}) replaces the stock WY belonging to Real Estate (Table~\ref{tabstage2}) in Stage 2; the stock EFX belonging to Industrials (Table~\ref{tabstage3a}) replaces the stock DD belonging to Materials (Table~\ref{tabstage3}) in Stage 3; the stock OKE belonging to Energy (Table~\ref{tabstage4a}) replaces the stock TROW belonging to Financials (Table~\ref{tabstage4}) in Stage 4; and two stocks MRO and FTI belonging to Energy (Table~\ref{tabstage5a}) replace stocks UNM and AFL belonging to Financials (Table~\ref{tabstage5}) in Stage 5. Overall, we notice tiny change in top $10$ influential stocks in each stage, hence the robustness of our framework.

\begin{table}[h]
\centering
\caption{Basic statistics of $5$ conserved networks based on the FDR correction.}\label{tabfdr}
 \begin{tabular}{|c|c|c|c|c|c|}
 \hline
    &\textbf{$|V|$} &\textbf{$|E|$} &\textbf{$\langle k\rangle$} &\textbf{$\langle c\rangle$} &\textbf{$\langle l\rangle$}\cr\hline
    Conserved Network 1 &422 &5545   &26.2796  &0.4589 &106.4866\cr\hline
    Conserved Network 2 &422 &49960  &236.7773 &0.7955 &9.3906\cr\hline
    Conserved Network 3 &422 &47412  &224.7014 &0.8092 &3.4824\cr\hline
    Conserved Network 4 &422 &68310  &323.7441 &0.8863 &1.2318\cr\hline
    Conserved Network 5 &422 &48849  &231.5118 &0.7994 &3.4502\cr\hline
 \end{tabular}
\end{table}

Then, we adopt the FDR for the multiple hypothesis test to relax the assumption of independence of the 
Bonferroni correction. As shown in Table~\ref{tabfdr}, the number of edges of the $5$ conserved networks are $3$ to $10$ times of those in Table~\ref{alpha1}. As a consequence, the networks are much dense. For example, conserved network 4, corresponding to the late transition from the crisis to the normal state, takes the value $1.2318$ of the average shortest path length. Intuitively, it is not the case of reality. In the stock market, the correlation matrix $\bm{W}$ always contain much noise, and therefore a restrictive procedure may perform better.

\begin{table}[h]
\centering
\caption{Top $10$ influential stocks identified by the PageRank algorithm from the conserved network corresponding to the normal stage before the 2008 crisis.}\label{tabpr}
 \begin{tabular}{|c|c|c|c|}
 \hline
    \textbf{Stock} &\textbf{Sector} &\textbf{$PR$} &\textbf{Cumulative return}\cr\hline
    PH  &Industrials            &0.0178 &1.6564\cr\hline
    RJF &Financials             &0.0123 &1.2366\cr\hline
    TFC &Financials             &0.0119 &0.9615\cr\hline
    HIG &Financials             &0.0103 &1.0541\cr\hline
    BEN &Financials             &0.0092 &1.3221\cr\hline
    WEC &Utilities              &0.0090 &1.1311\cr\hline
    GS  &Financials             &0.0088 &1.6365\cr\hline
    TER &Information Technology &0.0083 &0.9465\cr\hline
    SO  &Utilities              &0.0082 &1.0510\cr\hline
    MTB &Financials             &0.0078 &0.9397\cr\hline
 \end{tabular}
\end{table}

Finally, we employ the Google's PageRank algorithm~\cite{pagerank} to identify influential stocks. As a paradigmatic example of centrality-based ranking algorithm, PageRank has been found application in
a vast range of real systems, although it was devised originally to rank web pages. In Table~\ref{tabpr}, we show the results via the PageRank algorithm for the conserved network corresponding to the normal stage before the 2008 crisis. As aforesaid, stocks belonging to Real Estate should be more influential in this stage. However, none of the top $10$ influential stocks resulting from PageRank fall within that sector. We also observe contradictory results in other stages (not shown here).

\section{Conclusion}
The study of structure and dynamics of stock markets has attracted much attention from economists, mathematicians and physicists. An increasing interest is constructing reliable stock networks and analyzing their evolution~\cite{mastromatteo12jsm,squartini18pr}. Most approaches, however, have focused on the macro characteristics of the networks with less predictive power.

In this paper, we have addressed this issue from the micro point of view by characterizing spreading influence of each stock and identifying most influential stocks in a dynamic way. For this purpose, we first divided the investigating period of a stock market into a large number of windows through the moving-window method. Then, we constructed a stock network for each window by the significance test of stock correlations. Furthermore, we generated several conserved networks to extract various backbones of the market under different stages. Finally, we used order statistics to rank nodal influence and identified most influential stocks for each conserved network. To illustrate its efficacy, we have applied this procedure to stocks belonging to S\&P 500 Index from January 2006 to April 2010 and constructed $5$ conserved networks, based on which we identified various influential stocks under different stages and recovered the 2008 financial crisis from the evolutionary perspective.

We note, however, the stock market is too complex to be predicted. The present framework could be generalized by incorporating more physical and structural properties of the market. The comprehensive investigation of the market dynamics from both global and local aspects will be subjected to future research.

\section*{Acknowledgments}
We are grateful to referees for their valuable comments. This work was supported by the Natural Science Foundation of China under Grant Nos. 12071281 and 11771277.

\section*{Appendix. Results for $\alpha=0.02$}

\begin{table}[h]
\centering
\caption{Basic statistics of $5$ conserved networks for $\alpha=0.02$.}\label{alpha2}
 \begin{tabular}{|c|c|c|c|c|c|}
 \hline
    &\textbf{$|V|$} &\textbf{$|E|$} &\textbf{$\langle k\rangle$} &\textbf{$\langle c\rangle$} &\textbf{$\langle l\rangle$}\cr\hline
    Conserved Network 1 &422 &586   &2.7773   &0.1930 &410.3142\cr\hline
    Conserved Network 2 &422 &8611  &40.8104  &0.4771 &154.3755\cr\hline
    Conserved Network 3 &422 &10616  &50.3128  &0.5549 &130.0294\cr\hline
    Conserved Network 4 &422 &26478 &125.4882 &0.7035 &46.4160\cr\hline
    Conserved Network 5 &422 &8698  &41.2228  &0.4928 &177.7238\cr\hline
 \end{tabular}
\end{table}

\begin{table}[h]
\centering
\caption{Top $10$ influential stocks of the conserved network corresponding to the normal stage before the 2008 crisis.}\label{tabstage1a}
 \begin{tabular}{|c|c|c|c|}
 \hline
    \textbf{Stock} &\textbf{Sector} &\textbf{$Q$} &\textbf{Cumulative return}\cr\hline
    RJF &Financials  &0.9903 &1.2366\cr\hline
    ESS &Real Estate &0.9811 &1.2327\cr\hline
    FRT &Real Estate &0.9811 &1.4069\cr\hline
    KIM &Real Estate &0.9811 &1.3755\cr\hline
    BEN &Financials  &0.9531 &1.3221\cr\hline
    REG &Real Estate &0.9531 &1.2188\cr\hline
    BXP &Real Estate &0.9531 &1.3574\cr\hline
    UDR &Real Estate &0.9531 &1.0163\cr\hline
    DHR &Health Care &0.9468 &1.4872\cr\hline
    AVB &Real Estate &0.9442 &1.2967\cr\hline
 \end{tabular}
\end{table}

\begin{table}[h]
\centering
\caption{Top $10$ influential stocks of the conserved network corresponding to the early transition from the normal state to the crisis.}\label{tabstage2a}
 \begin{tabular}{|c|c|c|c|}
 \hline
    \textbf{Stock} &\textbf{Sector} &\textbf{$Q$} &\textbf{Cumulative return}\cr\hline
    AMP  &Financials             &1      &0.8696\cr\hline
    GL   &Financials             &1      &0.9815\cr\hline
    XOM  &Energy                 &0.9967 &0.9380\cr\hline
    GE   &Industrials            &0.9812 &0.9034\cr\hline
    BEN  &Financials             &0.9719 &0.7921\cr\hline
    MET  &Financials             &0.9717 &0.8784\cr\hline
    DD   &Materials              &0.9626 &0.8589\cr\hline
    AXP  &Financials             &0.9625 &0.7710\cr\hline
    TROW &Financials             &0.9443 &0.9508\cr\hline
    GPC  &Consumer Discretionary &0.9263 &0.7987\cr\hline
 \end{tabular}
\end{table}

\begin{table}[h]
\centering
\caption{Top $10$ influential stocks of the conserved network corresponding to the stage during the crisis.}\label{tabstage3a}
 \begin{tabular}{|c|c|c|c|}
 \hline
    \textbf{Stock} &\textbf{Sector} &\textbf{$Q$} &\textbf{Cumulative return}\cr\hline
    BEN  &Financials  &1      &0.5093\cr\hline
    TROW &Financials  &1      &0.5309\cr\hline
    PCAR &Industrials &0.9905 &0.5449\cr\hline
    AXP  &Financials  &0.9812 &0.2824\cr\hline
    ARE  &Real Estate &0.9719 &0.3619\cr\hline
    ITW  &Industrials &0.9625 &0.6188\cr\hline
    AMP  &Financials  &0.9625 &0.3576\cr\hline
    MMM  &Industrials &0.9572 &0.6185\cr\hline
    EFX  &Industrials &0.9533 &0.6871\cr\hline
    BXP  &Real Estate &0.9440 &0.3377\cr\hline
 \end{tabular}
\end{table}

\begin{table}[h]
\centering
\caption{Top $10$ influential stocks of the conserved network corresponding to the late transition from the crisis to the normal state.}\label{tabstage4a}
 \begin{tabular}{|c|c|c|c|}
 \hline
    \textbf{Stock} &\textbf{Sector} &\textbf{$Q$} &\textbf{Cumulative return}\cr\hline
    ETN  &Industrials &1      &1.6504\cr\hline
    PCAR &Industrials &0.9998 &1.5776\cr\hline
    EMR  &Industrials &0.9906 &1.4723\cr\hline
    DOV  &Industrials &0.9719 &1.5289\cr\hline
    MMM  &Industrials &0.9717 &1.5338\cr\hline
    RTX  &Industrials &0.9698 &1.4746\cr\hline
    SWK  &Industrials &0.9626 &1.4621\cr\hline
    CMI  &Industrials &0.9534 &1.8777\cr\hline
    WMB  &Energy      &0.9440 &1.6594\cr\hline
    OKE  &Energy      &0.9404 &1.6499\cr\hline
 \end{tabular}
\end{table}

\begin{table}[h]
\centering
\caption{Top $10$ influential stocks of the conserved network corresponding to the normal state after the crisis.}\label{tabstage5a}
 \begin{tabular}{|c|c|c|c|}
 \hline
    \textbf{Stock} &\textbf{Sector} &\textbf{$Q$} &\textbf{Cumulative return}\cr\hline
    L   &Financials  &1      &1.1066\cr\hline
    ALB &Materials   &0.9906 &1.2975\cr\hline
    DOV &Industrials &0.9810 &1.3617\cr\hline
    EMN &Materials   &0.9810 &1.2640\cr\hline
    WMB &Energy      &0.9809 &1.3315\cr\hline
    MRO &Energy      &0.9716 &0.9854\cr\hline
    HON &Industrials &0.9715 &1.2016\cr\hline
    VNO &Real Estate &0.9713 &1.2271\cr\hline
    EMR &Industrials &0.9618 &1.2794\cr\hline
    FTI &Energy      &0.9534 &1.2919\cr\hline
 \end{tabular}
\end{table}


\newpage

\section*{References}

\begin{thebibliography}{99}

\bibitem{mantegna00book}
Mantegna R N and Stanley H E 2000 \emph{An Introduction to Econophysics: Correlations and Complexity in Finance} (Cambridge: Cambridge University Press)

\bibitem{campbell97book}
Campbell J, Lo A W and MacKinlay A C 1997 \emph{The Econometrics of Financial Markets} (Princeton: Princeton University Press)

\bibitem{laloux9prl}
Laloux L, Cizeau P, Bouchaud J P and Potters M 1999 \emph{Phys. Rev. Lett.} \textbf{83} 1467

\bibitem{plerou99prl}
Plerou V, Gopikrishnan P, Rosenow B, Amaral L A N and Stanley H E 1999 \emph{Phys. Rev. Lett}. \textbf{83} 1471

\bibitem{potters21book}
Potters M and Bouchaud J P 2021 \emph{A First Course in Random Matrix Theory} (Cambridge: Cambridge University Press)

\bibitem{eom09phya}
Eom C, Oh G, Jung W S, Joeng H and Kim S 2009 \emph{Physica} A \textbf{388} 900

\bibitem{song11pre}
Song D M, Tumminello M, Zhou W X and Mantegna R N 2011 \emph{Phys. Rev.} E \textbf{84} 026108

\bibitem{jiang14sc}
Jiang X F, Chen T T and Zheng B 2014 \emph{Sci. Rep.} \textbf{4} 5321

\bibitem{wang17chaos}
Wang D, Zhang X, Horvatic D, Podobnik B and Stanley H E 2017 \emph{Chaos} \textbf{27} 023104

\bibitem{djauhari16jsm}
Djauhari M A and Gan S L 2016 \emph{J. Stat. Mech.} 093401

\bibitem{newman03book}
Newman M E J 2003 \emph{SIAM Rev.} \textbf{45} 167

\bibitem{bonanno03pre}
Bonanno G, Caldarelli G, Lillo F and Mantegna R N 2003 \emph{Phys. Rev.} E \textbf{68} 046130

\bibitem{onnela04epjb}
Onnela J P, Kaski K and Kertesz J 2004 \emph{Eur. Phys. J.} B \textbf{38} 353

\bibitem{tumminello05pnas}
Tumminello M, Aste T, Di Matteo T and Mantegna R N 2005 \emph{Proc. Natl. Acad. Sci. USA} \textbf{102} 10421

\bibitem{boginski05csda}
Boginski V, Butenko S and Pardalos P M 2005 \emph{Comput. Statist. Data Anal.} \textbf{48} 431

\bibitem{tumminello07epjb}
Tumminello M, Di Matteo T, Aste T and Mantegna R N 2007 \emph{Eur. Phys. J.} B \textbf{55} 209

\bibitem{liu11qf}
Liu J, Tse C K and He K 2011 \emph{Quant. Finance} \textbf{11} 817

\bibitem{wang17jeic}
Wang G J, Xie C and Chen S 2017 \emph{J. Econ. Interact. Coord.} \textbf{12} 561

\bibitem{xu18phya}
Xu X J, Wang K, Zhu L and Zhang L J 2018 \emph{Physica} A \textbf{509} 1080

\bibitem{wilinski13phya}
Wilinski M, Sienkiewicz A, Gubiec T, Kutner R and Struzik Z R 2013 \emph{Physica} A \textbf{392} 5963

\bibitem{wilinski18jsm}
Wilinski M, Ikeda Y and Aoyama H 2018 \emph{J. Stat. Mech} 023405

\bibitem{xu17sc}
Xu R, Wong W K, Chen G and Huang S 2017 \emph{Sci. Rep.} \textbf{7} 41379

\bibitem{musciotto18pc}
Musciotto F, Marotta L, Piilo J and Mantegna R N 2018 \emph{Palgrave Commun.} \textbf{4} 92

\bibitem{yang13mplb}
Yang C, Shen Y and Xia B Y 2013 \emph{Mod. Phys. Lett.} B \textbf{27} 1350022

\bibitem{bardoscia12jsm}
Bardoscia M, Livan G and Marsili M 2012 \emph{J. Stat. Mech.} P08017

\bibitem{bongiorno21epl}
Bongiorno C and Challet D 2021 \emph{EPL} \textbf{133} 48001

\bibitem{abhijit13sci}
Abhijit B, Chandrasekhar A G, Esther D and Jackson M O 2013 \emph{Science} \textbf{341} 1236498

\bibitem{wang16epjb}
Wang S, Sch\"{a}fer R and Guhr T 2016 \emph{Eur. Phys. J. B} \textbf{89} 105

\bibitem{benzaquen17jsm}
Benzaquen M, Mastromatteo I, Eisler Z and Bouchaud J P 2017 \emph{J. Stat. Mech.} P023406

\bibitem{roy11proc}
Roy R B and Sarkar U K 2011 \emph{Proc. of International Conference on Advances in Social Networks Analysis and Mining} (Kaohsiung, Taiwan) pp 567

\bibitem{fdr}
Benjamini Y and Hochberg Y 1995 \emph{J. Roy. Stat. Soc.} B \textbf{57} 449

\bibitem{onnela03pre}
Onnela J P, Chakraborti A, Kaski K, Kert\'{e}sz J and Kanto A 2003 \emph{Phys. Rev.} E \textbf{68} 056110

\bibitem{stiglitz16book}
Stiglitz J E 2016 \emph{Towards a General Theory of Deep Downturns} (Cambridge: Palgrave Macmillan)

\bibitem{freeman78sn}
Freeman L C 1978-1979 \emph{Soc. Netw.} \textbf{1} 215

\bibitem{bonacich87ajs}
Bonacich P F 1987 \emph{Am. J. Soc.} \textbf{92} 1170

\bibitem{bavelas48ho}
Bavelas A 1948 \emph{Hum. Organ.} \textbf{7} 16

\bibitem{freeman77soc}
Freeman L C 1977 \emph{Sociometry} \textbf{40} 35

\bibitem{das10book}
Dekking F M, Kraaikamp C, Lopuha\"{a} H P and Meester L E 2005 \emph{A Modern Introduction to Probability and Statistics} (London: Springer)

\bibitem{pagerank}
Langville A and Meyer C 2006 \emph{Google's PageRank and Beyond: The Science of Search Engine Rankings} (Princeton: Princeton University Press)

\bibitem{mastromatteo12jsm}
Mastromatteo I, Zarinelli E and Marsili M 2012 \emph{J. Stat. Mech.} P03011

\bibitem{squartini18pr}
Squartini T, Caldarelli G, Cimini G, Gabrielli A and Garlaschelli D 2018 \emph{Phys. Rep.} \textbf{757} 1

\end{thebibliography}
\end{document}